\documentclass[conference]{IEEEtran}
\IEEEoverridecommandlockouts

\makeatletter
\def\ps@IEEEtitlepagestyle{%
  \def\@oddfoot{\mycopyrightnotice}%
  \def\@oddhead{\hbox{}\@IEEEheaderstyle\leftmark\hfil\thepage}\relax
  \def\@evenhead{\@IEEEheaderstyle\thepage\hfil\leftmark\hbox{}}\relax
  \def\@evenfoot{}%
}
\def\mycopyrightnotice{%
  \begin{minipage}{\textwidth}
  \centering \scriptsize
  Copyright~\copyright~2024 \textbf{IEEE Engineering in Medicine and Biology Society (IEEE EMBC)}. Personal use of this material is permitted. Permission from IEEE must be obtained for all other uses, in any current or future media, including reprinting/republishing this material for advertising or promotional purposes, creating new collective works, for resale or redistribution to servers or lists, or reuse of any copyrighted component of this work in other works by sending a request to pubs-permissions@ieee.org.
  \end{minipage}
}
\makeatother

\usepackage{fancyhdr}

\usepackage{cite}
\usepackage{amsmath,amssymb,amsfonts}
\usepackage{xcolor}
\usepackage[vlined]{algorithm2e}

\SetCommentSty{mycommfont}

\usepackage{dblfloatfix}
\usepackage{booktabs} 
\usepackage{makecell} 
\usepackage{xcolor} 
\usepackage{colortbl} 

\usepackage{graphicx}
\usepackage{textcomp}

\usepackage{amsmath,amssymb}
\usepackage{makecell}
\usepackage{multirow}
\usepackage{caption}
\usepackage{float}

\usepackage{capt-of}
\usepackage{booktabs}
\usepackage{varwidth}
\usepackage{amsmath}
\usepackage{algorithmic}
\usepackage{setspace, etoolbox}
\usepackage{hyperref}

\usepackage{amssymb}
\usepackage{makecell}
\usepackage{cite}
\usepackage{amsmath,amssymb,amsfonts}

\usepackage{textcomp}

\def\BibTeX{{\rm B\kern-.05em{\sc i\kern-.025em b}\kern-.08em
    T\kern-.1667em\lower.7ex\hbox{E}\kern-.125emX}}

\makeatletter
    \newcommand{\linebreakand}{%
      \end{@IEEEauthorhalign}
      \hfill\mbox{}\par
      \mbox{}\hfill\begin{@IEEEauthorhalign}
    }
\makeatother

\makeatletter
\newcommand{\newlineauthors}{%
  \end{@IEEEauthorhalign}\hfill\mbox{}\par
  \mbox{}\hfill\begin{@IEEEauthorhalign}
}
\makeatother

\begin{document}

\title{Bayesian-Guided Generation of Synthetic Microbiomes with Minimized Pathogenicity}

\author{\IEEEauthorblockN{Nisha Pillai}
\IEEEauthorblockA{\textit{Mississippi State University}\\
pillai@cse.msstate.edu}
\and
\IEEEauthorblockN{Bindu Nanduri}
\IEEEauthorblockA{\textit{Mississippi State University}\\bnanduri@cvm.msstate.edu}

\and
\IEEEauthorblockN{Michael J Rothrock Jr.}
\IEEEauthorblockA{\textit{USDA-ARS}\\michael.rothrock@usda.gov}

\newlineauthors
\IEEEauthorblockN{Zhiqian Chen}
\IEEEauthorblockA{\textit{Mississippi State University}\\zchen@cse.msstate.edu}
\and
\IEEEauthorblockN{Mahalingam Ramkumar}
\IEEEauthorblockA{\textit{Mississippi State University}\\ramkumar@cse.msstate.edu}
}

\maketitle

\begin{abstract}

Synthetic microbiomes offer new possibilities for modulating microbiota, to address the barriers in multidtug resistance (MDR) research. We present a Bayesian optimization approach to enable efficient searching over the space of synthetic microbiome variants to identify candidates predictive of reduced MDR. Microbiome datasets were encoded into a low-dimensional latent space using autoencoders. Sampling from this space allowed generation of synthetic microbiome signatures. Bayesian optimization was then implemented to select variants for biological screening to maximize identification of designs with restricted MDR pathogens based on minimal samples. Four acquisition functions were evaluated: expected improvement, upper confidence bound, Thompson sampling, and probability of improvement. Based on each strategy, synthetic samples were prioritized according to their MDR detection. Expected improvement, upper confidence bound, and probability of improvement consistently produced synthetic microbiome candidates with significantly fewer searches than Thompson sampling. By combining deep latent space mapping and Bayesian learning for efficient guided screening, this study demonstrated the feasibility of creating bespoke synthetic microbiomes with customized MDR profiles.

\end{abstract}


\section{Introduction} \label{introduction}

Multi-drug resistance (MDR) is a growing concern in the treatment of various infectious diseases. By identifying which parts of the microbiome most significantly influence MDR, researchers can develop targeted strategies to combat this resistance. This could involve manipulating the microbiome to enhance the presence of beneficial microbes that naturally suppress drug-resistant strains~\cite{relman2018microbiome}. Further research into the development of synthetic microbiomes that mitigate antibiotic resistance is therefore crucial~\cite{kogut2019effect}.

Synthetic microbiome samples provide researchers with a powerful tool for gaining a deeper understanding of microbial communities. This allows them to identify how changes in the microbiome affect health, enable disease states, and increase multidrug resistance. Synthetic microbiome samples accelerate microbiology research and advance microbiome-related medical challenges in a faster, more focused manner. In this study, we are proposing an architecture to generate relative abundance values for synthetic microbiome samples capable of minimizing MDR. The architecture we propose involves selecting samples based on diversity to strengthen a classification network, creating an efficient synthetic sample from lower dimension random data using an autoencoder methodology, and employing Bayesian optimization to determine the synthetic microbiome sample predictive of reduced MDR within fewer iterations.

\begin{figure*}[h]
\centering
\includegraphics[width=0.95\textwidth]{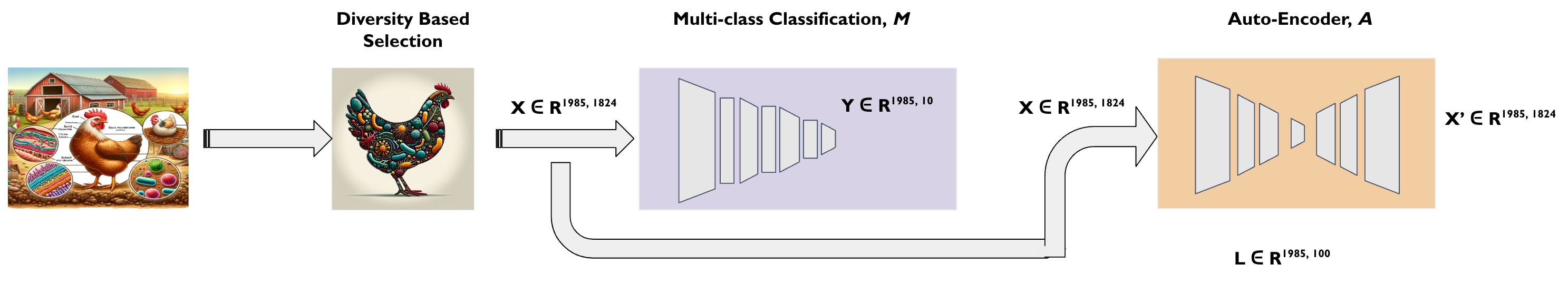}
\caption{The learning phase of synthetic data generation architecture. 
}
\label{fig:design_diagram1}
\end{figure*}

Our architectural design comprises two distinct segments. The initial segment (see figure~\ref{fig:design_diagram1}) involves constructing a multi-class classification network. Within this framework, our focus is on enhancing the learning efficiency of classification. Additionally, we establish an autoencoder generative model. This model is adept at representing microbiome samples in a compressed, low-dimensional space, and is capable of reconstructing these samples with minimal quality degradation. In the subsequent segment (see figure~\ref{fig:design_diagram2}), we employ a strategy that involves generating samples with a Gaussian distribution in the latent space. Utilizing the decoder component of the autoencoder, we then create synthetic microbiome samples. This synthesis is achieved more efficiently through the application of a Bayesian optimization technique, allowing for the creation of these samples in reduced number of steps.

\section{Dataset} \label{dataset}

Data used in this study are described in detail in \cite{hwang2020farm, pillai2022ensemble, rothrock2019microbiomic}. In this section, we provide a summary of these materials. Subsection~\ref{mdr} discusses how \textit{Salmonella}, \textit{Listeria}, and \textit{Campylobacter} multidrug resistance (target variables) was determined. Subsection \ref{microbiome} describes the methods used to determine microbiome samples and relative abundance values (input variables). Data from eleven pasture-raised broiler farms in the southeastern United States with flock sizes ranging between 25-1500 birds was used in this study. Preharvest (Feces and Soil) and postharvest (Ceca, WCR-F, and WCR-P) samples were included in this study.

The dataset consists of 1985 samples and 1824 operational taxonomic units (OTUs) representing the microbiomes present. For each sample, the relative abundance values of the 1824 OTUs are included in the dataset. This allows for a comprehensive analysis of the microbial community composition across the 1985 samples.

\subsection{Cultural Isolation for Antibiotic Sensitivity Testing} \label{mdr}

\textit{Salmonella}, \textit{Campylobacter}, and \textit{Listeria} isolates were cultured and tested for antibiotic resistance using standard NARMS (\url{www.cdc.gov/narms}) protocols. The isolates in the dataset exhibit varying levels of antibiotic resistance, ranging between 0 to 9. Any isolates that were found to be resistant to three or more antibiotics were classified as multidrug resistant (MDR). This categorization of multidrug resistance provides important insight into the prevalence and patterns of antibiotic resistance within the bacterial population represented by the isolates. Detailed methods for testing antibiotic sensitivity in each of the three bacterial species were provided in~\cite{hwang2020farm, pillai2023towards,pillai2024towards, gireesan2024deep}, including the specific antibiotics and concentration ranges tested, as well as the incubation conditions and quality control strains used.

\subsection{Microbiome Analysis} \label{microbiome}

DNA was extracted from samples using a semi-automated hybrid protocol combining enzymatic and mechanical methods~\cite{rothrock2014hybrid}. The DNA concentration was determined spectrophotometrically after purification. The V4 domain of the bacterial 16S rRNA gene was amplified and sequenced using the Illumina MiSeq platform~\cite{caporaso2011global}. The QIIME v1.9.1 pipeline was used to process the raw sequence reads~\cite{caporaso2010qiime}, including chimera checking (\url{http://drive5.com/uchime/gold.fa}), clustering into operational taxonomic units (OTUs), taxonomic assignment~\cite{desantis2006greengenes}, sequence alignment (PyNAST~\cite{caporaso2010pynast}), and phylogenetic tree generation. This analysis workflow allowed for characterization of the microbial communities present in the samples.

\begin{figure*}[tp]
\centering
\includegraphics[width=0.9\textwidth]{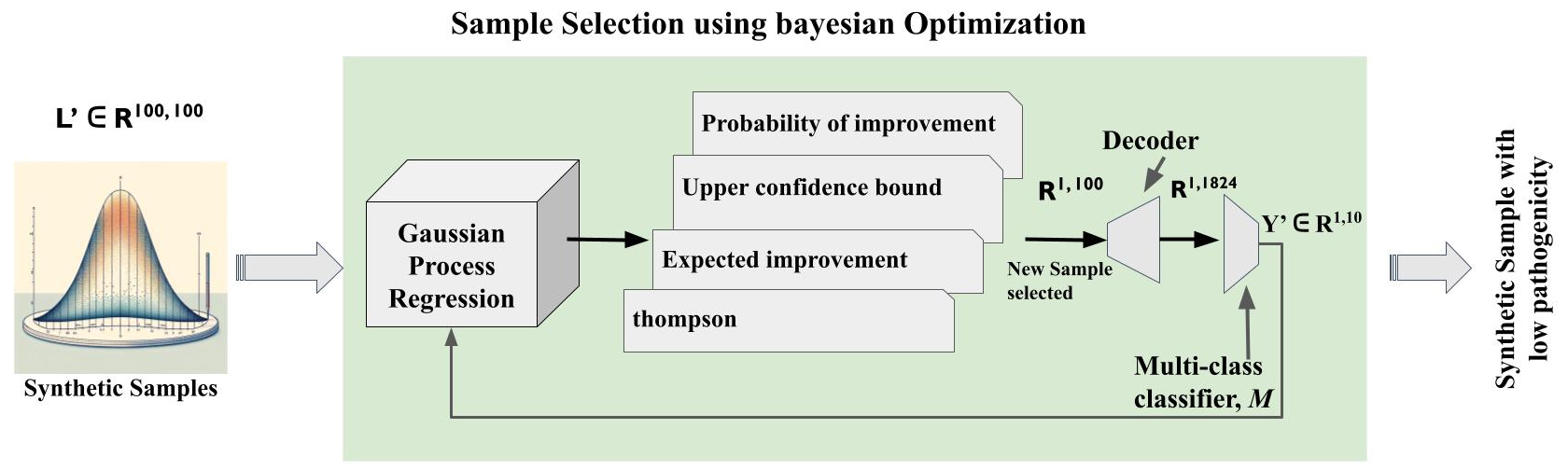}
\caption{Generation of synthetic microbiomes using Bayesian optimization in few iterations.}
\label{fig:design_diagram2}
\end{figure*}

\section{Approach} \label{approach}
\subsection{Multi-Class Classification} \label{classification}
Training an accurate multi-class classification neural network typically requires large, balanced datasets. However, collecting ample biological data can be challenging, making it difficult to train an efficient model when resources are scarce. To mitigate class imbalance, we employ an oversampling technique. We first randomly oversample the minority class to increase its representation. Next, we apply Synthetic Minority Over-sampling Technique (SMOTE)~\cite{chawla2002smote} to further balance the classes. This method includes a combination of random over-sampling as well as SMOTE over-sampling technique. Using these augmented microbiome samples, $ X \in R^{N, 1824}$, we train a neural network to predict multi-class target $ Y \in R^{N, 10}$, values ranging from 0 to 9 (see figure \ref{fig:design_diagram1}). Here, $N$ is the number of microbiome samples. Oversampling enables us to train an accurate model despite limited and imbalanced biological data.

\subsection{Diverse Point Selection} \label{diverse_selection}

Training neural networks on diverse data is vital for developing accurate, fair, and robust models that generalize well. Incorporating diversity in the training set provides multiple key benefits, most notably enhancing model performance, generalization, and representation learning. Exposing the model to varied data points during training augments its ability to extrapolate insights and patterns. This equips the model to adapt effectively when deployed in new environments, unseen scenarios, and with different subgroups. Overall, diversity enables the model to transcend the specifics of the training data. This results in flexible, broadly capable neural networks suited for real-world usage across a spectrum of situations and populations.

To select diverse microbiome samples, we employ a determinantal point process (DPP)~\cite{kulesza2012determinantal, kulesza2011k, kulesza2011learning}. As a probability distribution over all possible subsets, a DPP promotes diversity by assigning higher probability to more varied subsets. Specifically, the likelihood of sampling a subset is proportional to the determinant of a kernel matrix generated from elements within that subset. This kernel encapsulates similarity between items in the ground set, making dissimilar items more likely to co-occur. Consequently, DPPs preferentially pick diverse, representative subsets where entries are not excessively redundant. By modeling subset diversity and sampling accordingly, DPPs enable the tailored selection of microbial samples to emphasize heterogeneity. This generates varied samples suited to learning tasks requiring broad coverage over the microbial spectrum. The determinantal formulation innately encourages the discovery of novel and complementary subsets from the microbiome.

\subsection{Auto-Encoder} \label{auto-encoder} 
Autoencoders are an unsupervised artificial neural network technique useful for learning representations of input data in an efficient, compressed latent space. The goal of an autoencoder is to reproduce its inputs - it takes an input, encodes it into a lower-dimensional code, and tries to reconstruct the original input from this code. Auto-Encoder, $A$ (see figure~\ref{fig:design_diagram1}), consists of an encoder model that compresses the microbiome input into a latent code ($L$), and a decoder model that decompresses this encoding back into the original input space($X'$). A bottleneck in the network forces it to capture the most salient features of the data. The autoencoder is trained to minimize the difference between the microbiome input and reconstructed microbiome output.

\begin{figure*}[tp]
\centering
\includegraphics[width=0.9\textwidth]{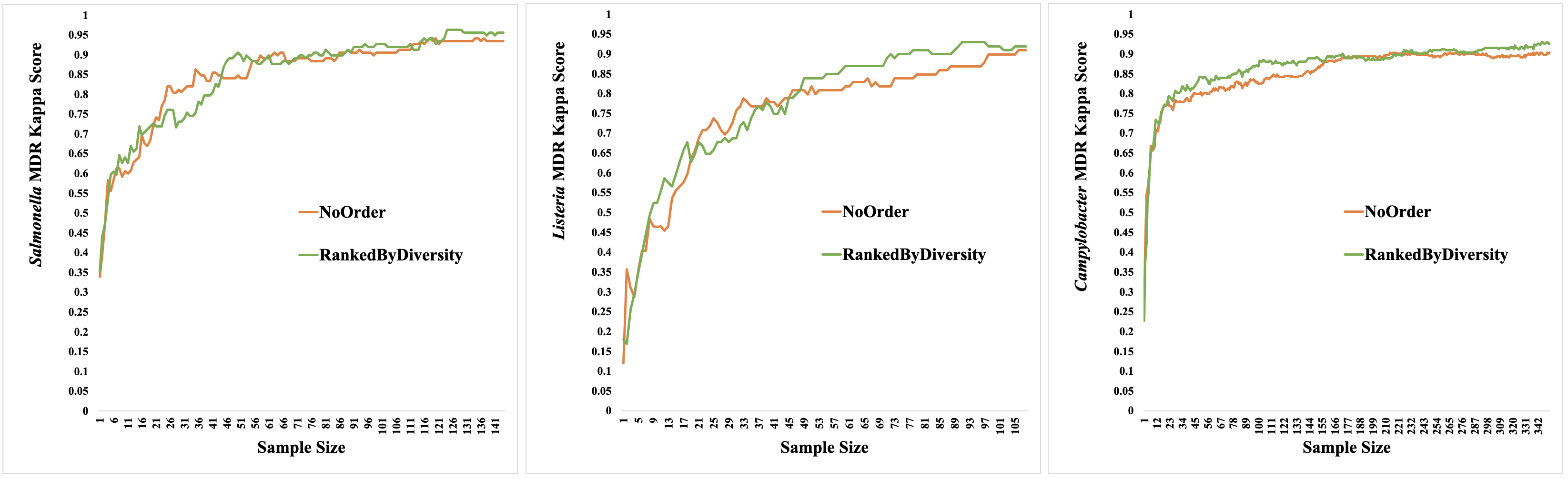}
\caption{Multi-class classification performance comparison at every iteration for \textit{Salmonella}, \textit{Listeria}, and \textit{Campylobacter} MDR.}
\label{fig:comparison_nn}
\end{figure*} 

\subsection{Random Data Generation in Latent Space} \label{latent_random_data}

In the second phase of our architecture (see figure~\ref{fig:design_diagram2}), we generate synthetic microbiome samples from the latent space to find ones predicting low prevalence of MDR food-borne  pathogens . We produce 100 sample latent variable datasets two ways: 1) Sampling a Gaussian matching the latent distribution mean and standard deviation of original microbiome dataset $X$;
 2) Applying Latin hypercube sampling~\cite{mckay1979comparison} for evenly-spaced random data. Gaussian sampling assumes the data follows a normal or Gaussian distribution, which occurs commonly in many real-world datasets. This makes it a reasonable default assumption in the absence of other information. Latin hypercube sampling ensures full coverage over the entire range of each input variable with fewer samples compared to simple random sampling. The samples are spread evenly across each dimension’s distribution. This space-filling property provides better representation of the full variability of the input space. Also the stratified spatial coverage leads to unbiased Monte Carlo estimates of the output variables. Errors converge faster without the need for extremely large sample sizes. Evaluating across these two different distributions tests the capability to reliably regenerate benign microbial communities from any latent sample source. Ultimately, validating efficacy across differing synthetic distributions enables the reliable discovery of microbiome compositions minimizing likelihood of multidrug resistance. This leverages the latent space to safely and effectively explore and identify candidate samples conferring antibiotic resistance without risk of exposure.

\subsection{Bayesian Optimization} \label{bayesian_optimization}

Bayesian optimization~\cite{movckus1975bayesian} is an efficient method for optimizing unknown black-box functions that are expensive to evaluate. Here, we apply Bayesian optimization to find microbiome samples with minimal presence of MDR food-borne pathogens like \textit{Salmonella}, \textit{Listeria}, and \textit{Campylobacter}. We construct a multi-label probabilistic Gaussian process regression model that predicts presence of MDR \textit{Salmonella}, \textit{Listeria}, and \textit{Campylobacter} from 0 to 1 based on a microbiome sample’s features. By modeling uncertainty, an acquisition function decides which new candidate sample to generate next using a latent space decoder model of auto-encoder, $A$ in order to gather the most useful information about the optimum. After calculating each new candidate’s predicted pathogen presence using multi-class classification model, $M$, the Gaussian process model is updated and the sampling iterates until convergence towards microbiome samples with lowest pathogen presence. We examine four acquisition functions commonly used in Bayesian optimization: Firstly, Thompson sampling is used in Bayesian optimization as a method of selecting the next point to be evaluated. It maintains a probabilistic model (posterior distribution) of the objective function based on the data observed so far. Using this posterior distribution, it samples candidate solutions and evaluates the candidate that looks most promising. In the second acquisition function, probability of improvement (pi) is used to calculate the probability that evaluating a point will lead to an improvement over the current best observed value. Third, we use an upper confidence bound (UCB), which aims to balance exploitation by maximizing selection around the predicted mean and exploration by selecting areas with high uncertainty. By explicitly and cleanly combining both exploitation (predicted mean) and exploration (uncertainty), UCB offers an efficient, scalable, and effective acquisition function for Bayesian optimization. We use the expected improvement (EI) as the fourth acquisition function to figure out where to sample next in latent samples. The key idea behind expected improvement is to calculate the expected (mean) improvement over the current best objective value if we were to evaluate at a candidate point. EI quantifies both the improvement that could be achieved over current best objective value and the likelihood of achieving said improvement based on the predicted mean and uncertainty at candidate point. Bayesian optimization is well-suited for optimizing microbiome engineering due to few required evaluations for optimal solutions and seamless trade-off between exploration and exploitation of existing knowledge. In this study, we intend to generate a synthetic microbiome sample that can predict low levels of \textit{Salmonella}, \textit{Listeria}, and \textit{Campylobacter} MDR within fewer iterations.

\section{Experiments And Results} \label{experiments}

\begin{table}[bp]

\renewcommand{\arraystretch}{1.75}
 \centering
  \scalebox{1}{
  \begin{tabular}{lcc}
    \toprule

& \thead{No Order} & \thead{Rank By Diversity}\\

    \midrule
\textbf{\textit{Salmonella} MDR}  &	0.93490 & \textbf{0.95660}   \\ 
\textbf{\textit{Listeria} MDR} & 0.90929 & \textbf{0.91940} \\ 
\textbf{\textit{Campylobacter} MDR} &  	0.91560 & \textbf{0.93606}  \\

  \bottomrule
\end{tabular}}
 \caption{Comparison of multi-classification accuracy (model with randomly sampled train data vs ranked by diversity) with test data.}
  \label{tab:prediction_metrics}

\end{table}

A study of 1985 pastured poultry samples collected from eleven pastured farms in the southeast of the United States at various stages bird growth, harvesting and processing was conducted. The presence of antibiotic resistance was not detected in approximately 75\% of the samples with \textit{Campylobacter}. There is, however, a resistance to one or more antibiotics in all Listeria samples. As compared to \textit{Listeria} and \textit{Campylobacter}, the number of \textit{Salmonella} samples with antibiotic resistance is very limited. To evaluate the model performance and assess potential overfitting, we calculated the classification accuracy and Kappa score over both the test dataset and the dataset generated by an autoencoder. Comparing the model’s performance on these different datasets provides an estimate of the degree of overfitting, if any, in the model.

\begin{table}[bp]
\centering
\renewcommand{\arraystretch}{1.5}

\scalebox{0.9}{
\begin{tabular}{lcccc}
\toprule
& \multicolumn{2}{c}{\thead{Accuracy}} & \multicolumn{2}{c}{\thead{Kappa Score}} \\
\cmidrule(lr){2-3} \cmidrule(lr){4-5}
& \thead{No Order} & \thead{Rank By\\ Diversity} & \thead{No Order} & \thead{Rank By\\ Diversity} \\
\midrule
\textbf{\textit{Salmonella} MDR} & 0.8037 & \textbf{0.8971} & 0.7748 & \textbf{0.8820} \\
\textbf{\textit{Listeria} MDR} & 0.8083 & \textbf{0.8166} & 0.7684 & \textbf{0.7780} \\
\textbf{\textit{Campylobacter} MDR} & 0.8235 & \textbf{0.9002} & 0.7983 & \textbf{0.8859} \\
\bottomrule
\end{tabular}
}
\caption{Comparison of multi-classification performance (model with randomly sampled train data vs ranked by diversity) with auto-encoder recreated test data.}
\label{tab:vae_prediction_metrics}
\end{table}

\paragraph{Multi-Class Classification} We have hidden layers with sizes of 1000, 500, and 100 in our multi-classification neural network. Based on the MDR profiles of each pathogen, the output layer units are selected. We used the ReLU~\cite{agarap2018deep} function as a non-linear activation function in the hidden layers, followed by a softmax layer. We implemented our classification system using scikit-learn (version 0.24.2)~\cite{sklearn_api} and Pytorch (version 2.0.0)~\cite{NEURIPS2019_9015}. The prediction performance of a multi-class classification network built with normally shuffled training samples and ranked by diversity was examined. Since it is a multi-class classification, both accuracy and kappa score are calculated. A comparison of the accuracy of the two methods is provided in Table~\ref{tab:prediction_metrics}. We noted that the performance of the prediction is enhanced when training samples are re-ordered according to diversity rank. The kappa score of prediction is compared with each interval of training samples in the same way. Figure~\ref{fig:comparison_nn} illustrates the incremental improvement in performance at each step when training samples are ranked according to diversity. It follows that the order of training samples influences the training accuracy, and for better model performance, it is recommended that the training set be resampled according to diversity.

\begin{table*}[tp]
\renewcommand{\arraystretch}{0.85}
 \centering
 \scalebox{0.9}{
  \begin{tabular}{ll|cccccccccccc|}
      \toprule
 \multicolumn{14}{l|}{\textbf{\textit{Salmonella}}}\\ \hline
\toprule
 & & \multicolumn{12}{c|}{Iterations}\\
Sampling & BOMethod & 1	& 100 &	200 &	283	&290&	300&	400&	500&	574&	600	&630&	631\\ \hline
\midrule

 \multirow{4}{*}{\makecell[c]{Latin-\\Hypercube} } & thompson & 0.98	& 4.1E-5	& 0.03 &	\textcolor{red}{4.4E-6} & & &&&&&&\\

& ei & 0.01 &	7E-3 &	1E-4 & & & 8E-4 &	1E-4 &	1E-3 & & 0.3	& & 	\textcolor{red}{4.4E-6}\\ 

 & ucb & 0.01 & 	7E-3&	5.7E-5 & & & 5E-4 &	1E-4 &	0.01 & & 2E-3 & 	\textcolor{red}{4.4E-6} & \\

& pi & 0.03 &	7E-3 &	1E-4 & & & 8.2E-5	 &2E-3 & 1E-3 & & 0.3	 & &\textcolor{red}{4.4E-6}\\ \hline

\multirow{4}{*}{\makecell[c]{Gaussian} } & thompson & 1E-4 & 	1E-3 &	1E-4 & & \textcolor{red}{4.4E-6} & &&&&&&\\

& ei & 0.8 &	1E-4 & 	1E-4 & & & 5E-4	& 1E-4 & 0.01 & & 2E-2	&\textcolor{red}{4.4E-6}&  \\ 

 & ucb & 1E-4 &	7E-4 &	1E-4 & & & 8E-4 & 1E-4 & 1E-4 & & 0.3	& & \textcolor{red}{4.4E-6} \\

& pi & 0.3 &7E-3&1E-4 & & & 8E-4& 1E-4 &	0.01 & & 2E-3 &	\textcolor{red}{4.4E-6} &  \\ \hline




\\
\multicolumn{2}{l|}{\textbf{\textit{Listeria}}}& 1&100&	181	&182&	183	&200&	300&	386&	400	&500&	517&	579\\ \hline
\midrule

 \multirow{4}{*}{\makecell[c]{Latin-\\Hypercube} } & thompson & 1E-3 & 7E-2 & & & & 1E-2 & 1E-3 & & 6.5E-5 &	2E-4 & & \textcolor{red}{1E-15}\\

& ei & 6E-2	&3E-3	& & &	\textcolor{red}{1E-15}&&&&&&&\\ 

 & ucb & 0.1 &	3E-3	& & \textcolor{red}{1E-15} &&&&&&&&\\

& pi & 4.4E-5 &	3E-3	& & & 	\textcolor{red}{1E-15}&&&&&&&\\ \hline

\multirow{4}{*}{\makecell[c]{Gaussian} } & thompson & 3E-2 & 2E-3 & & & &  1E-3 & 5E-3 &	\textcolor{red}{1E-15} &&&&\\

& ei & 5E-3 & 1E-2 &	\textcolor{red}{1E-15} &&&&&&&&& \\ 

 & ucb & 1E-3 &	3E-3	& & &	\textcolor{red}{1E-15} &&&&&&&\\

& pi & 5E-4 & 3E-3	& & &	\textcolor{red}{1E-15} &&&&&&&\\ \hline





\\
\multicolumn{2}{l|}{\textbf{\textit{Campylobacter}}} & 1	&100	&181	&182&	183&	200	&300&	386	&400	&500	&517	&579\\ \hline
\midrule

 \multirow{4}{*}{\makecell[c]{Latin-\\Hypercube} } & thompson & 0.8	&0.9 & & & & 0.6	&0.1 & & 0.8&0.5	& & \textcolor{red}{2E-3}\\

& ei & 0.9 &	0.1	& & & \textcolor{red}{2E-3}&&&&&&&\\ 

 & ucb &0.2 & 	0.1	& & \textcolor{red}{2E-3} &&&&&&&&\\

& pi & 0.8 &	0.1	& & & 	\textcolor{red}{2E-3}&&&&&&&\\ \hline

\multirow{4}{*}{\makecell[c]{Gaussian} } & thompson & 0.9 &	0.9 &&&&0.8 &	0.8 &	\textcolor{red}{2E-3} &&&&\\

& ei & 0.2 &	0.4 &	\textcolor{red}{2E-3}  &&&&&&&&&\\ 

 & ucb & 0.8 &	0.1	& & & \textcolor{red}{2E-3}&&&&&&&\\

& pi & 0.2 &	0.1	& & & \textcolor{red}{2E-3} &&&&&&&\\ \hline





\end{tabular}}
 \caption{Bayesian optimization acquisition functions are compared with two random data generation methods to determine the fastest selection procedure for finding synthetic samples with the lowest levels of MDR predictions.}
\label{tab:comparison_bayesian}

\end{table*}

\begin{figure*}[t]
\centering
\includegraphics[width=0.9\textwidth]{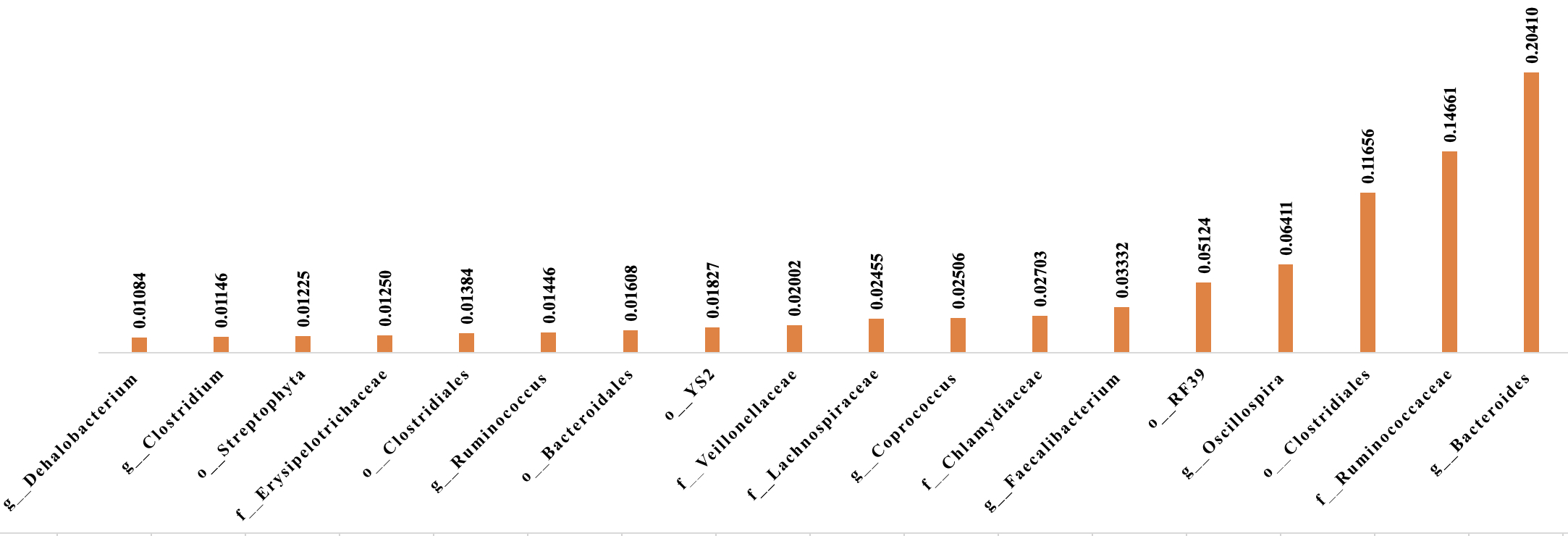}
\caption{Top influencing microbiomes and their relative abundance values within a synthetic microbiome sample selected by our framework that predicts low MDR.}
\label{fig:microbiome_selection}
\end{figure*}

\paragraph{Auto-Encoder} We used a deep neural network for encoder and decoder models. The encoder consists of hidden sizes 1800, 1600, 1400, 1200, 1000, 800, 600, 400, and 200, while the latent representation dimension is 100. The decoder used the same number of hidden units as the encoder in reverse order. At every layer, ReLU activation functions were used as nonlinear functions. The decoder ends with 1824 output units and a sigmoid activation function. The reconstruction loss is used in an auto-encoder model for backpropagation. The auto-encoder framework is built using Keras neural network library~\cite{chollet2015keras}. Based on testing with several hidden unit combinations, the above combination proved to be the most effective. To evaluate the quality of reconstruction, we verified the classification accuracy using reconstructed data. The efficiency of diversity was also evaluated by comparing the training done in normal order and ranked by diversity. Table~\ref{tab:vae_prediction_metrics} shows that training samples when ranked by diversity performed better than when ranked by normal order in all cases. The Kappa score and accuracy are calculated in both cases, and observed improved performance in both cases.

\paragraph{Bayesian Optimization} To determine which function works best based on our data, we evaluated two random data generation methods with four Bayesian optimization acquisition functions. For \textit{Salmonella}, \textit{Listeria}, and \textit{Campylobacter} MDR, table \ref{tab:comparison_bayesian} provides comparisons of the number of iterations (samples taken from an unclassified pool) required to attain the lowest MDR prediction using each Bayesian optimization method based on random data generation. The \textit{Salmonella} dataset is limited and the sample counts for each class are highly disproportionate. Thompson sampling method seems to provide the best result for such datasets, allowing for the selection of the synthetic sample with the least number of iterations from both data sequences. In the case of \textit{Listeria}, and \textit{Campylobacter}, which have comparatively higher balanced data and counts, the expected improvement, probability of improvement, and upper confidence bounds seem to favor selecting the synthetic sample that predicts lowest MDR score in few iterations. It is evident from this that all three approaches can be applied to such a situation.

\paragraph{Synthetic Microbiome Selection} Figure~\ref{fig:microbiome_selection} illustrates the synthetic microbiome top features and its relative abundance values chosen by our model as the one producing the lowest \textit{Salmonella}, \textit{Listeria}, and \textit{Campylobacter} MDR levels. There is ongoing research~\cite{khan2020gut, tan2019investigations} into the use of probiotics to promote a healthy balance of gut microbiota, including beneficial \textit{Bacteroides} species~\cite{ty2022performance}, as a way to improve poultry health and productivity. \textit{Bacteroides} species, the most abundant in poultry help in breaking down complex carbohydrates, proteins, and lipids, facilitating nutrient absorption, important for the efficient utilization of feed. Members of the \textit{Ruminococcaceae} family are known for their ability to degrade complex plant polysaccharides. In poultry, this is crucial for the breakdown of dietary fibers, contributing to more efficient nutrient absorption and digestion. Based on our evaluations, \textit{Ruminococcaceae} family has been identified as the second most influential microbiome responsible for the reduction of pathogens. Interestingly, the research~\cite{diaz2018tannins} indicate that members of the order \textit{Clostridiales}, predominantly those belonging to the \textit{Ruminococcaceae} family, may play a role in the productivity of birds. It is interesting to note that our results indicate that order \textit{Clostridiales} have the 3rd highest influence on reduced pathogen prevalence. The \textit{Ruminococcaceae} family is a group of bacteria that is part of the \textit{Firmicutes} phylum. According to our results, the majority of the influential microbiomes are found in the \textit{Firmicutes} phylum in the lowest pathogen formula. Even though the \textit{Chlamydiaceae} family has a relatively low relative abundance value, we have found that it plays an important role in our findings. A zoonotic disease, avian chlamydiosis, is caused by \textit{Chlamydiaceae}~\cite{ornelas2020cross,li2017chlamydia}, so the findings warrant further investigation.

\section{Related Research} \label{related}

The use of synthetic data is gaining an increasingly prominent role in data and machine learning workflows in healthcare and biomedical research~\cite{achuthan2022leveraging, walonoski2020synthea}. Synthetic data enables researchers to build more robust models and conduct analyses with greater statistical power than possible with real-world datasets alone. The purpose of this study is to create synthetic microbiome samples that can predict reduced MDR in poultry. Synthetic data generation using autoencoders has emerged as a popular and effective approach in biomedical research. Autoencoders are neural networks that compress input data into a latent space representation and then reconstruct the outputs. By training autoencoders on real biomedical datasets, researchers can learn robust lower dimensional encodings capturing the most salient properties and patterns in complex data. Researchers~\cite{titus2018unsupervised} used variational autoencoders (VAEs) as an unsupervised deep learning approach to model DNA methylation patterns across breast cancer tumors. The method learns latent representations capturing complex relationships in tumor epigenetic profiles without the need for manually labeled data. Similarly, researchers~\cite{pratella2021survey} have demonstrated the utility of autoencoders and variational autoencoders, for modeling diverse biomedical data types in both disease and health contexts.

Latent Space Bayesian Optimization (LSBO)~\cite{boyar2023latent} approaches integrates two key machine learning techniques - variational autoencoders (VAEs) and Bayesian optimization (BO) - to enable targeted optimization and generation of novel data points. Stanton et.al.~\cite{stanton2022accelerating} developed a framework combining a denoising autoencoder with a multi-task Gaussian process model to optimize the design of novel fluorescent proteins. This approach allies the latent space representation learning of autoencoders with the sample-efficient optimization of Gaussian processes. Similarly, we have employed Bayesian optimization at the latent space level to generate synthetic microbiome samples to predict low levels of food-borne pathogens.

\section{Conclusion} \label{conclusion}

In conclusion, our research has made significant strides in exploring the potential of synthetic microbiome samples for predicting phenotypes of interest, notably multi-drug resistance (MDR). By introducing an efficient and diversified multi-class classification method, we have substantially enhanced our capabilities in pathogen prediction. Furthermore, the implementation of an autoencoder framework has opened avenues for generating synthetic samples. These samples demonstrate reduced pathogenicity, an advancement that could have substantial implications in the field of microbiome research and pathogen management. Additionally, our adoption of a Bayesian approach has streamlined the iteration process, allowing for more efficient progression in our research. Through these contributions, we have strengthened the overall pipeline for synthesizing, analyzing, and refining synthetic microbiome data for phenotype prediction. The validity of our results has been thoroughly tested against reasonable baselines, ensuring the robustness of our findings. Moreover, our investigation has identified key microbiome contributors influencing the studied phenotypes. These findings significant in their own right are also corroborated by supporting research, providing a deeper understanding of the microbiome’s role in pathogen behavior and resistance patterns. Our multi-faceted framework shows promising capability to elucidate and predict microbial community-level phenotypes relevant to food safety, human health and disease.

\section*{Funding}

Dataset used in this study is provided by the Agricultural Research Service, USDA CRIS Project ``Reduction of Invasive Salmonella enterica in Poultry through Genomics, Phenomics and Field Investigations of Small MultiSpecies Farm Environments'' \#6040-32000-011-00-D. This research was supported by the Agricultural Research Service, USDA NACA project entitled ``Advancing Agricultural Research through High Performance Computing'' \#58-0200-0-002 and 58-6064-3-017.

\bibliography{reference}

\end{document}